\documentclass[]{achemso}

\usepackage{textcomp}
\usepackage{amsmath}    
\usepackage{amssymb}    
\usepackage{siunitx}    
\usepackage{graphicx}   
\usepackage{csquotes}
\makeatletter
\let\acs@movepunct\relax
\makeatother

\author{Zhijia Zhang}
\affiliation[University of Manchester]
{Department of Physics and Astronomy, University of Manchester, Manchester, UK}
\altaffiliation{National Graphene Institute, University of Manchester, Manchester, UK.}
\author{Mohsen Moazzami Gudarzi} 
\affiliation[University of Manchester]
{Department of Physics and Astronomy, University of Manchester, Manchester, UK}
\altaffiliation{National Graphene Institute, University of Manchester, Manchester, UK.}
\email{mohsen.moazzamigudarzi@manchester.ac.uk}
\author{Jiatong Mao} 
\affiliation[University of Manchester]
{Department of Physics and Astronomy, University of Manchester, Manchester, UK}
\altaffiliation{National Graphene Institute, University of Manchester, Manchester, UK.}
\author{Ziwei Wang} 
\affiliation[University of Manchester]
{Department of Physics and Astronomy, University of Manchester, Manchester, UK}
\altaffiliation{National Graphene Institute, University of Manchester, Manchester, UK.}
\author{Zakhar Bedran} 
\affiliation[University of Manchester]
{Department of Physics and Astronomy, University of Manchester, Manchester, UK}
\author{Chuhongxu Chen} 
\affiliation[University of Manchester]
{Department of Physics and Astronomy, University of Manchester, Manchester, UK}
\author{Milad Nonahal} 
\affiliation[University of Manchester]
{Department of Chemistry, University of Manchester, Manchester, UK}
\author{Ivan Timokhin} 
\affiliation[University of Manchester]
{Department of Physics and Astronomy, University of Manchester, Manchester, UK}
\author{Artem Mishchenko }
\email{artem.mishchenko@manchester.ac.uk}
\affiliation[University of Manchester]
{Department of Physics and Astronomy, University of Manchester, Manchester, UK}
\altaffiliation{National Graphene Institute, University of Manchester, Manchester, UK.}
\author{Qian Yang}
\email{qian.yang@manchester.ac.uk}
\affiliation[University of Manchester]
{Department of Physics and Astronomy, University of Manchester, Manchester, UK}
\altaffiliation{National Graphene Institute, University of Manchester, Manchester, UK.}

\title{Rapid fabrication of clean van der Waals nanochannels using Mask and Stack method}

\begin{document}

\begin{abstract}
Two-dimensional (2D) nanochannels have emerged as a pivotal platform for exploring nanoscale hydrodynamics and electrokinetics. Conventional fabrication methods to make nanochannels often introduce polymer contamination and require lengthy processing, limiting device performance and scalability. Here we introduce the Mask $\&$ Stack method, employing silicon nitride stencil mask combined with dry transfer stacking to rapidly fabricate ultraclean vdW nanochannels within hours. This polymer-free approach preserves pristine interfaces, confirmed by atomic force microscopy and Raman spectroscopy, and yields nanochannel devices exhibiting reproducible ionic transport and long-term stability. The streamlined process is compatible with diverse 2D materials and promising for upscale production. Our method advances the fabrication of nanofluidic and 2D heterostructure devices, facilitating applications in quantum transport, photonics, energy harvesting, and sensing technologies requiring high-throughput, contamination-free heterostructure architectures.

\end{abstract}

\section{Introduction}

Nanocapillary plays a central role in biological systems as well as numerous technological sectors\cite{hille2001ionic,bocquet2010nanofluidics}. In living cells, the rapid transport of water and the highly selective passage of ions and proteins through nanocapillaries are essential for maintaining the delicate balance of life. Inspired by these natural systems, artificial nanofluidic platforms have been developed across diverse architectures, including bottom-up synthesis of carbon nanotubes, assembly of van der Waals (vdW) 2D nanocapillaries and top-down fabrication by electron irradiation or pipette pulling\cite{geim2021exploring,li2025carbon,zhu2019ion}. Although still in an early stage compared to their biological counterparts, artificial nanochannels are rapidly proving their potential in selective molecular transport, bio-sensing, energy harvesting, and neuromorphic computing\cite{ro2025iontronics}. 

In particular, 2D vdW nanochannels have greatly facilitated the experimental development in nanofluidics in recent years and led to new physics down to the atomic scale, including molecular and ionic transport that deviates from classic descriptions\cite{esfandiar2017size,radha2016molecular,gopinadhan2019complete}, anomalous behavior of water under confinement\cite{yang2020capillary,fumagalli2018anomalously,wang2025plane}, highly efficient electrokinetic phenomena\cite{emmerich2022enhanced} and memristor behavior promising for neuromorphic computing\cite{robin2023long,xiao2024neural}. All these can only be explored theoretically before the introduction of 2D nanochannels. Beyond nanofluidics, 2D nanochannels also bear considerable similarities with their biological counterparts, could help understand the fast transport and high selectivity mechanism in biological channels, and even reverse-engineering biological systems for improved diagnosis and therapy\cite{welch2021advances}. 

However, fabricating vdW nanochannels remains technically demanding despite detailed protocols\cite{bhardwaj2024fabrication}. A key challenge is contamination of the spacer surfaces by polymer resist residues introduced during electron beam lithography (EBL) and exacerbated by crosslinking in subsequent reactive ion etching (RIE). To mitigate this, laborious cleaning cycles -- typically involving solvent rinsing with acetone and isopropanol, followed by high-temperature annealing -- are repeatedly employed, yet residues often persist. Additional wet transfer steps for stacking further introduce contamination and extend the fabrication timeline to several weeks. These lengthy procedures, coupled with low device yield, have significantly hindered the broader development and application of 2D nanochannel technologies.

Here, we present a fast and robust fabrication strategy that overcomes these limitations by combining stencil lithography with all-dry vdW assembly -- the Mask $\&$ Stack method. Stencil mask has been widely used in microfabrication for etching and deposition processes with sub-micron resolution, demonstrated excellent reproducibility and scalability\cite{yong2016rapid,verschueren2018lithography,du2017stencil, zeng2025direct,barcons2022engineering}. Pre-patterned silicon nitride (SiN) stencils act as reusable etching masks for defining spacer layers, which are then stacked to form sandwiched nanochannels via dry transfer within hours. Because there is no interface exposure to polymers, this method, therefore, preserves the pristine cleanliness of 2D materials. The method is also broadly compatible with diverse 2D materials, enabling applications beyond nanofluidics in heterostructure physics, cavity-enhanced spectroscopy, and photonic architectures. Looking ahead, integration with robotic and high-throughput platforms offers a promising route toward scalable manufacturing of ultraclean nanofluidic devices and industrial adoption of vdW-based technologies.

\section{Results and Discussion}
\subsection{Rapid fabrication of vdW Nanochannels}
Below, we describe the fabrication process for vdW 2D nanochannels using the Mask $\&$ Stack method, as illustrated in \textbf{Figure 1a and Supplementary Video}. Wafer-scale freestanding SiN membrane production is first made using photolithography and EBL, followed by RIE processes, following the standard procedures for creating suspended SiN structures (see more details in \textbf{Methods, SiN spacers stencil fabrication}, and Supplementary Information \textbf{Figure S1}). A single 4-inch SiN/Si/SiN wafer typically yields over 60 stencils masks with designated patterns (\textbf{Figure S1b}), each mask can be reliably reused multiple times. For nanochannel device fabrication, a few or up to tens of parallel lines are patterned into the freestanding SiN membrane to define the spacer layer masks, as shown in \textbf{Figure S1c,d} and \textbf{Figure S2}. 

Once the mask is prepared, 2D crystals are placed on top of the mask, followed by backside RIE to form the spacer layer (\textbf{Figure 1a}, Step 1). Comparing to the conventional lithography-based spacer fabrication, this process removes entirely the polymer spin coating onto the spacer and maintains its pristine surface. The pristine spacer surface enables further vdW stacking through a dry transfer process (more details in \textbf{Mothods, Dry transfer method for vdW assembly}), instead of wet transfer that introduces extensive polymer on the spacer surface. To do this, a large 2D crystal exfoliated directly onto a PDMS stamp, serving as the top layer, is first used to pick up the spacer layer from the mask (Step 2), followed by the pickup of a bottom crystal (Step 3) to complete the sandwiched nanochannel assembly. The top crystal is deliberately chosen to be larger than the spacer layer for ease of the dry transfer process; this also ensures that, once the encapsulation is complete, the entire channel area remains shielded from airborne contaminants. The resulting sandwiched stack is then released onto a SiN membrane with a rectangular slit opening, such that the edge of the bottom layer spans the slit while the spacers align perpendicular to its long axis (Step 4), thereby enabling mass transport during measurements. Both steps 3 and 4 are carried out at elevated temperatures of $\sim$150°C, and vdW pressure helps expel airborne hydrocarbons and maintain clean interfaces\cite{jain2018minimizing}. The entire stacking process can be completed within hours, a dramatic improvement over conventional wet transfer techniques, which often require days up to weeks\cite{bhardwaj2024fabrication}. The shortened fabrication timeline also reduces device exposure to airborne hydrocarbons, thereby preserving the atomically smooth surface of 2D materials. 

Just before measurements, the fully encapsulated channels are opened using RIE to allow mass transport (\textbf{Figure 1b)}. The channel length is defined by photolithography or using the edge of another dry-transferred 2D materials as the etching mask, depending on the specific measurements requirements. Importantly, the encapsulated spacers remain clean even after polymer exposure during the lithography process. The final device can therefore be measured directly, without annealing at high temperatures (typically at 300/400°C), avoiding any thermal degradation of the channel walls. 

A wide range of 2D materials can be patterned this way because of the high etching resistivity of SiN using the RIE than for most 2D materials. It also enables patterning of hard-to-process materials using the conventional EBL method, such as thick graphite and gold crystals (\textbf{Figure S3 }and \textbf{S4}), where polymer resists wear off or cross link during the following RIE procedure and introduces complication to the fabrication process. This enables the fabrication of vdW heterostructures for nanoelectronic and nanophotonic applications.
\begin{figure}
    \centering
    \includegraphics[width=1\linewidth]{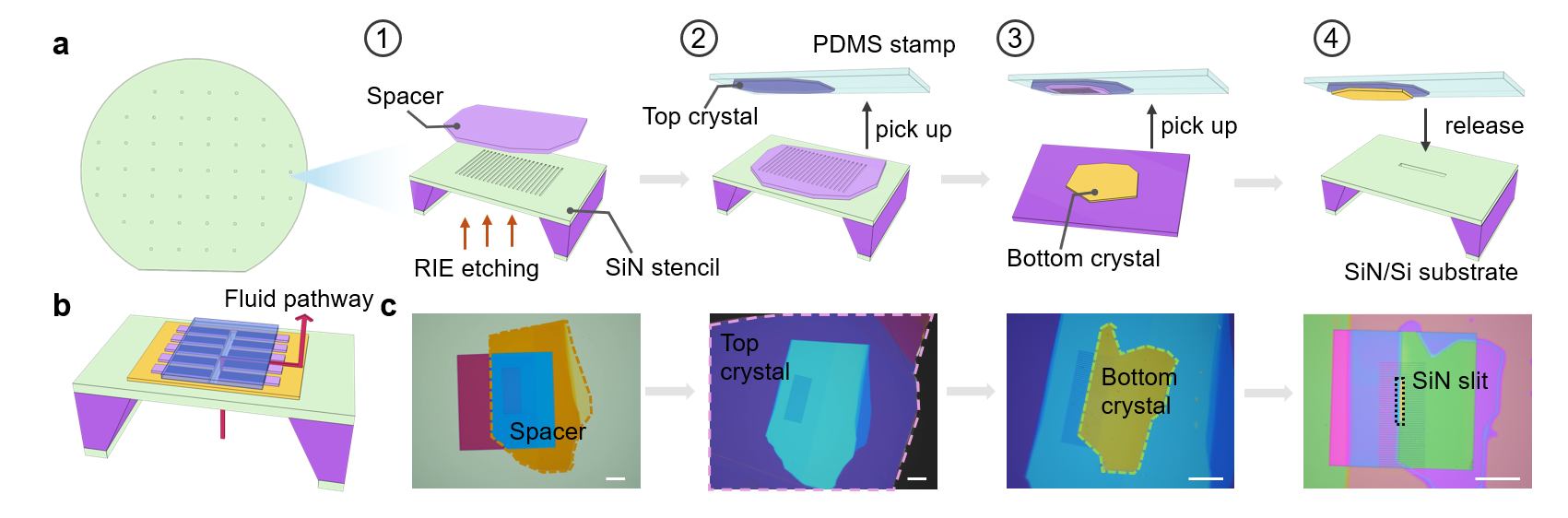}
    \caption{\textbf{Fabrication of vdW  nanochannel devices using SiN stencil mask.} \textbf{a}, Schematic illustration of the device fabrication workflow. Wafer-scale SiN mask production creating stencil masks with designated patterns, following standard lithography and etching processes. (Step 1) using pre-patterned SiN as etching mask, spacer layer is made through RIE etching from backside. (Step 2) the spacer layer is picked up by PDMS stamp using dry transfer method, with exfoliated top crystal on the stamp. (Step 3) a bottom crystal is then picked up to complete the sandwich stacking. (Step 4) the encapsulated nanochannel is aligned and released onto a SiN membrane with predefined slit opening. \textbf{b}, Schematic of the final vdW nanochannel device, the red arrow indicates the mass transport pathway through the nanochannels. \textbf{c}, Optical images from each fabrication step corresponding to the schematics shown in steps 1-4. Dashed lines show the contours of the named layers on each image. Scale bars, 20 $\mu$m.
}
    \label{Schem}
\end{figure}
\subsection{Ultra-clean nanochannels}
Surface contamination and polymer residues are known to significantly influence interfacial properties in nanofluidic systems\cite{yang2022effect,emmerich2024nanofluidics}. In the nanochannel architecture, spacers fabricated using the conventional EBL often retain substantial polymer residues on their surfaces and edges, which demands laborious cleansing procedures (\textbf{Figure S5}). These residues can obstruct channels or introduce inaccuracies in the effective channel height measurement once the devices are assembled, therefore require extensive cleaning to remove. Meanwhile, contamination at the interlayers will reduce the adhesion between the spacer and the top and bottom crystals, compromising the long-term stability of the devices \cite{sajja2021hydrocarbon}. This would lead to delamination of the sandwiched stacks during measurements, especially in solution and/or under high applied pressure. 

To demonstrate the cleanliness of the fabricated nanochannels, we first performed atomic force microscopy (AFM) measurements (\textbf{Methods, AFM measurements}). Here, we choose MoS$_2$ with a different number of layers, from monolayer to five layers, as the spacers. After back etching using SiN mask, these spacers were picked up by a top crystal (few layer graphene (FLG) or MoS$_2$) using PDMS stamp (Step 2). AFM measurements were carried out directly on the flipped spacers, supported on the top crystal and the PDMS stamp. 
\begin{figure}
    \centering
    \includegraphics[width=1\linewidth]{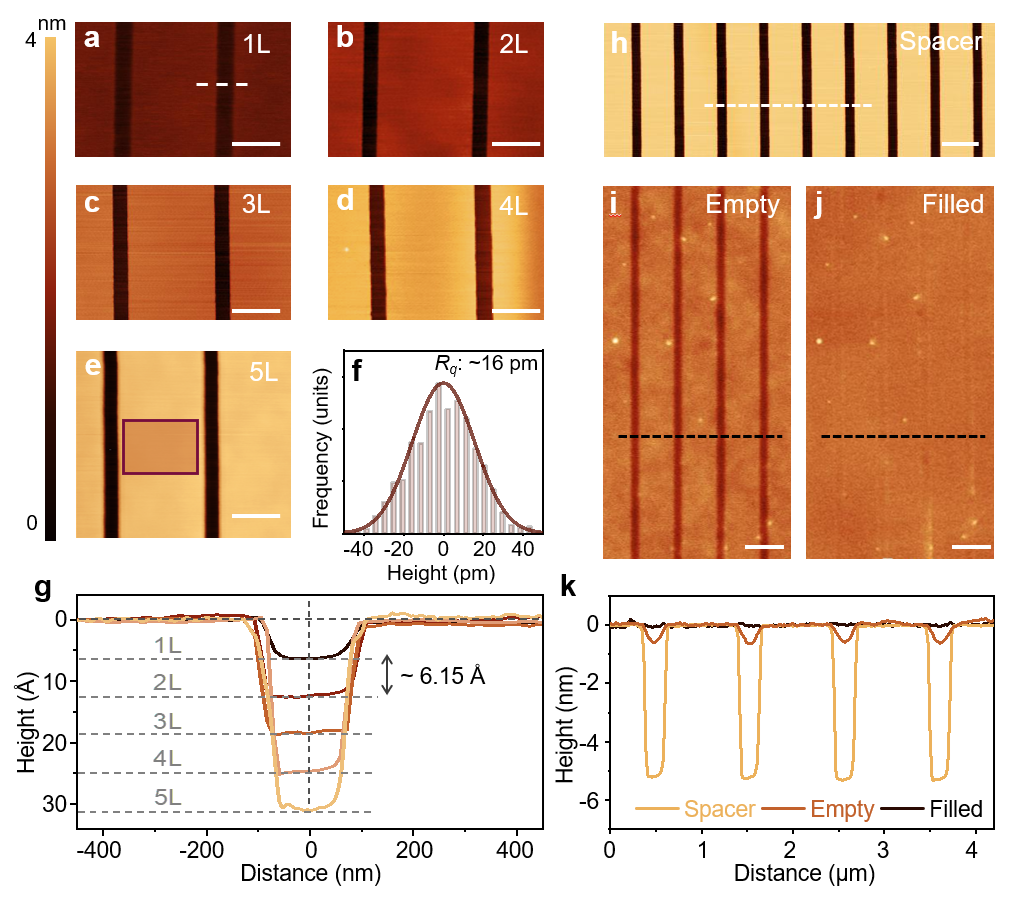}
    \caption{\textbf{Cleanliness of the nanochannels characterized by AFM.} \textbf{a–e}. AFM images of MoS$_2$ spacers with varying thicknesses: \textbf{a}, 1L, \textbf{b}, 2L, \textbf{c}, 3L, \textbf{d}, 4L, and \textbf{e}, 5L. These spacers are imaged on top of another 2D crystal and PDMS stamp. All spacers exhibit well-defined edges and clean surface. The dashed white line in \textbf{a} indicates the positions where the height profiles of these channels are taken.  The red box in \textbf{e} shows the area where the roughness in g is obtained. \textbf{f}, Histogram representation of surface roughness for the boxed region in \textbf{e}, showing RMS roughness of $\sim$16 pm. \textbf{g}, Height profiles across one of the slits in each AFM images in \textbf{a-e}, showing step height consistent with the reported interlayer spacing of MoS$_2$.  \textbf{h}, AFM image of a FLG spacer on graphite bottom crystal on PDMS stamp. \textbf{i}, AFM topography of the FLG  channel surface after assembly, showing visible sagging of the top layer above the channels. \textbf{j}, AFM topography of the same channels after filling with water, showing channels leveling up with the nearby spacers. \textbf{k}, Height profiles across spacers and channels as indicated by the white  dashed line in \textbf{h} and black dashed lines in \textbf{i} and \textbf{j}. Scale bars on all images: 1 $\mu$m.}
    \label{AFM}
\end{figure}
\textbf{Figure 2a-e } shows the AFM images of MoS$_2$ spacers with increasing number of layers. The AFM results show clean surfaces and well-defined channels, with edges of the channels sharp and free of contaminants, indicating minimal chemical and mechanical damage during the RIE and the following dry-transfer process. Our analysis shows line edge roughness  as low as 1 nm with correlation length exceeding 90 nm (\textbf{Figure S2}). Height profiles taken across the spacer (\textbf{Figure 2 a-e}) show distinct steps (\textbf{Figure 2g}), corresponding to the interlayer spacing of MoS$_2$ ($\sim$6.15 Å)\cite{murray1979thermal}. This further confirms the clean interface between the spacer and the underlying crystal after dry transfer. \textbf{Figure 2f} shows the root-mean-square (RMS) roughness of the transferred MoS$_2$ spacers (the area in the red box in panel \textbf{e}) on PDMS stamp, as low as $\sim$16 pm, approaching the instrumental noise limit\cite{lui2009ultraflat}. This is much smaller than the surface roughness of spacers fabricated using the conventional EBL method, shown in \textbf{Figure S5c}, which appear clean with no visible polymer residues, yet statistical analysis shows much higher surface roughness of around 216 pm. Reduced contamination and surface roughness influence the friction and surface charge density of the channels, factors that strongly influence the mass transport and fluid dynamics behaviors inside the nanochannels\cite{emmerich2022enhanced, qin2022study}. 

In addition, we show the uniformity of 40 monolayer MoS$_2$ spacers after etching reusing the same SiN mask, in \textbf{Figure S6}. AFM images of these channels exhibit a clean surface, well-defined edges and consistent widths, with statistical analysis yielding an average width of $151.0 \pm 5.5$\,nm, matching the width of the SiN mask used. These results confirm the advantages of our fabrication approach in making nanochannels for measurements and applications that require both ultraclean and high precision features.

Beyond surface cleanliness, we also show that the channel interior remains clean after assembly using the capillary filling method. Sagging of the top layer is often observed for nanochannels with relatively thin top, due to the vdW adhesion between the top layer and the channel sidewalls\cite{yang2020capillary}. When water fills inside channels, it effectively screens the adhesion which leads to the lift up of the sagged top layer. We show in \textbf{Figure 2h-k} the AFM images and height profiles of a nanochannel device with spacer height $\sim$5nm (yellow profile in Figure 2k). After cover the spacer with top crystal and complete the sandwiched assembly, the channels have initial sagging of $\sim$5.5 Å (AFM image in \textbf{Figure 2i} and the brown height profile marked as Empty in \textbf{Figure 2k}), which then level up to the nearby spacers once filled with water, as shown in \textbf{Figure 2j} and the black height profile in \textbf{Figure 2k}. We observed uniform sagging of the top layer throughout the 12 $\mu$m-long channels, indicating absence of contamination, as trapped hydrocarbons would form blister-like bulges and uneven channel height profiles along channels. We also show that all 11 channels in the $12\ \mu\mathrm{m} \times 12\ \mu\mathrm{m}$ imaged area display uniform height profiles, \textbf{Figure S7}), demonstrating excellent homogeneity of the channels fabricated using the Mask $\&$ Stack method.

\begin{figure}
    \centering
    \includegraphics[width=1\linewidth]{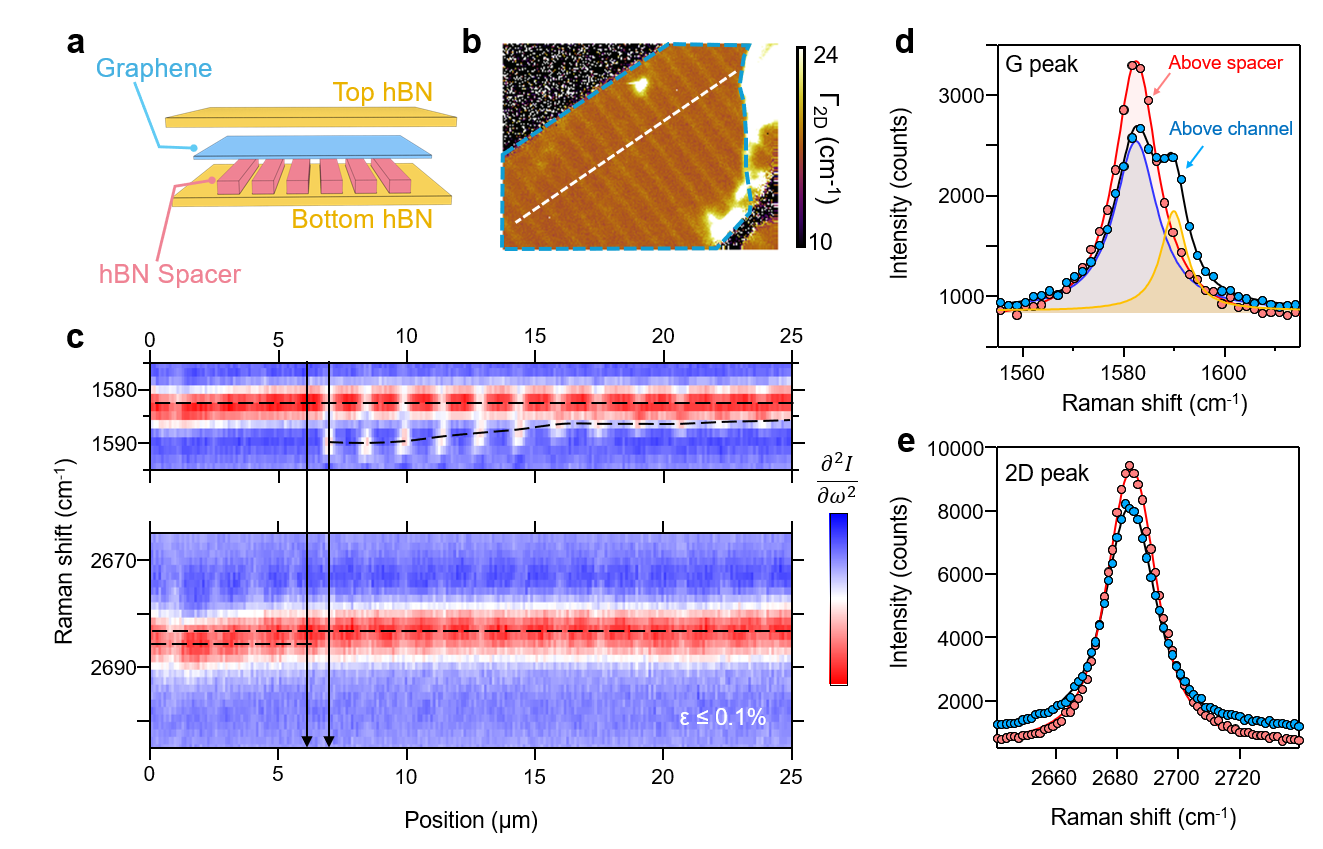}
    \caption{\textbf{Raman mapping of nanochannels}. \textbf{a}, Schematic illustration of the device cross-section, consisting of MLG encapsulated between the hBN spacer and top layer. \textbf{b}, Raman mapping of the FWHM of MLG 2D band $\Gamma_\mathrm{2D}$. The dashed blue line contours the MLG region. \textbf{c}, Second-derivative maps for the G (top) and 2D (bottom) peaks along the dashed white line in \textbf{b}. The horizontal dashed lines are visual guides for average peak positions. The two vertical solid lines mark representative positions for spacer and channel regions, respectively, used in \textbf{d} and \textbf{e}. \textbf{d-e}, Lorentzian fits to the G peak (\textbf{d}) and 2D peak (\textbf{e}) for the spacer area (red) and channel area (blue).}
    \label{Raman}
\end{figure}

To further characterize the cleanliness of the fabricated nanochannels, we employed Raman spectroscopy (\textbf{Methods, Raman Spectroscopy}). To this end, we embedded monolayer graphene (MLG) between the top hBN crystal and hBN spacer layers to enhance the spectral response (as illustrated in \textbf{Figure 3a}), taking advantage of the highly sensitive and well-calibrated MLG spectra to the presence of strain, doping, and structural disorder\cite{lee2012optical,neumann2015raman,pimenta2007studying}. Optical and AFM images of the device are shown in \textbf{Figure S8a} and \textbf{b}. We performed Raman hyperspectral mapping of the MLG-embedded nanochannel device, as shown in \textbf{Figure 3b} and \textbf{Figure S8c, d}. We observe alternating parallel features that correspond to the channel (narrow, light yellow) and spacer (wide, brown) regions for the MLG-embedded area, as contoured by the blue dashed line in \textbf{Figure 3b}. A line-scan taken across the channels, the white dashed line in \textbf{Figure 3b}, is presented in \textbf{Figure 3c}. Here we show the second derivative of G and 2D band intensity to enhance spectral contrast. The position of the G peak exhibits gradual shifts that correlate with the nanochannel geometry, suggesting local strain modulation due to sagging of MLG in the alternating channel-spacer geometry. In contrast, the 2D peak position remains largely uniform, indicating that the overall pristine quality of the graphene is preserved throughout the device. 

We further extracted the representative Raman spectra above the channel and spacer, as marked by vertical solid lines in \textbf{Figure 3c}, respectively. As shown in \textbf{Figure 3d}, Raman spectrum collected above the spacers shows a symmetric G peak, with full width at half maximum (FWHM, $\Gamma_\mathrm{G}$) 10.9 cm$^{-1}$. In contrast, we observed splitting of the G band in the area above channels, which we attribute to uniaxial strain that comes from sagging of MLG above channel areas together with top hBN and a local doping of graphene  \cite{liu2021split,PhysRevB.79.205433}. A more detailed analysis of spatially resolved strain and doping distributions is provided in \textbf{Supplementary Information, Raman analysis}. Note that the laser spot size (\(2\lambda/\pi N.A.\), about 380 nm in our case) is more than twice larger than the nanochannels' width (about 150 nm)\cite{cai2010thermal}. As a result, the spectrum over nanochannels represent a combination of both suspended (above channel) and supported (above spacer) regions, making it challenging to decouple the doping and strain in the suspended region. However, the 2D peaks shown in \textbf{Figure 3e} are nearly identical for both position and linewidth of the 2D band, indicating that the graphene lattice remains structurally uniform and free of additional disorder after nanochannel fabrication. The minimal differences in 2D band width between the two regions confirm that the fabrication process preserves the intrinsic quality of the graphene, with negligible impact from strain or doping. Overall, the Raman analysis supports the structural integrity and cleanliness of the vdW interfaces in our nanochannel devices. This method can be further established to monitor the local strain and doping environments of the nanochannels due to external stimuli.

The clean channels also exhibit excellent ion transport performances under various measurement conditions, with improved reproducibility and prolonged device lifetime. Below, through ionic transport measurement, we demonstrate that nanochannels fabricated using the Mask $\&$ Stack method serve as a robust and scalable platform for nanofluidic exploration. 

\subsection{Ionic Transport and Stability}
We fabricated nanochannel devices with monolayer MoS$_2$ spacers (\textbf{Figure 4a}) to demonstrate the nanofluidic transport and device stability. These channels represent an extreme confinement scenario ($\sim$6.15 Å channel height), with the size of the channel smaller than the hydrated diameter of the tested ions. Details about the ionic transport measurements can be found in \textbf{Methods}. As shown in \textbf{Figure 4b}, the measured ionic conductance exhibits typical surface-governed behavior, significantly exceeding the bulk conductance value and becoming saturated at low concentrations\cite{stein2004surface}. Conductance data fitted using the variable surface charge model (purple dashed curve) indicate the role of surface charge regulation under nanoconfined conditions\cite{secchi2016scaling}. The model yields a low surface charge density of \SI{100}{\micro\coulomb\per\square\meter} for the MoS$_2$ channel, which is attributed to the exceptional cleanliness of our nanochannels. Ionic transport measurements conducted using cations with increasing hydrated diameter reveal effective steric exclusion due to the size of the channel. That is, cations with larger hydrated diameter show much suppressed conductivity  through the channels (\textbf{Figure 4c}). This agrees with previous reports that ions with hydrated diameter larger than that of the channel size are expected to pass through the channel with reduced mobility \cite{esfandiar2017size}.

\begin{figure}
    \centering
    \includegraphics[width=1\linewidth]{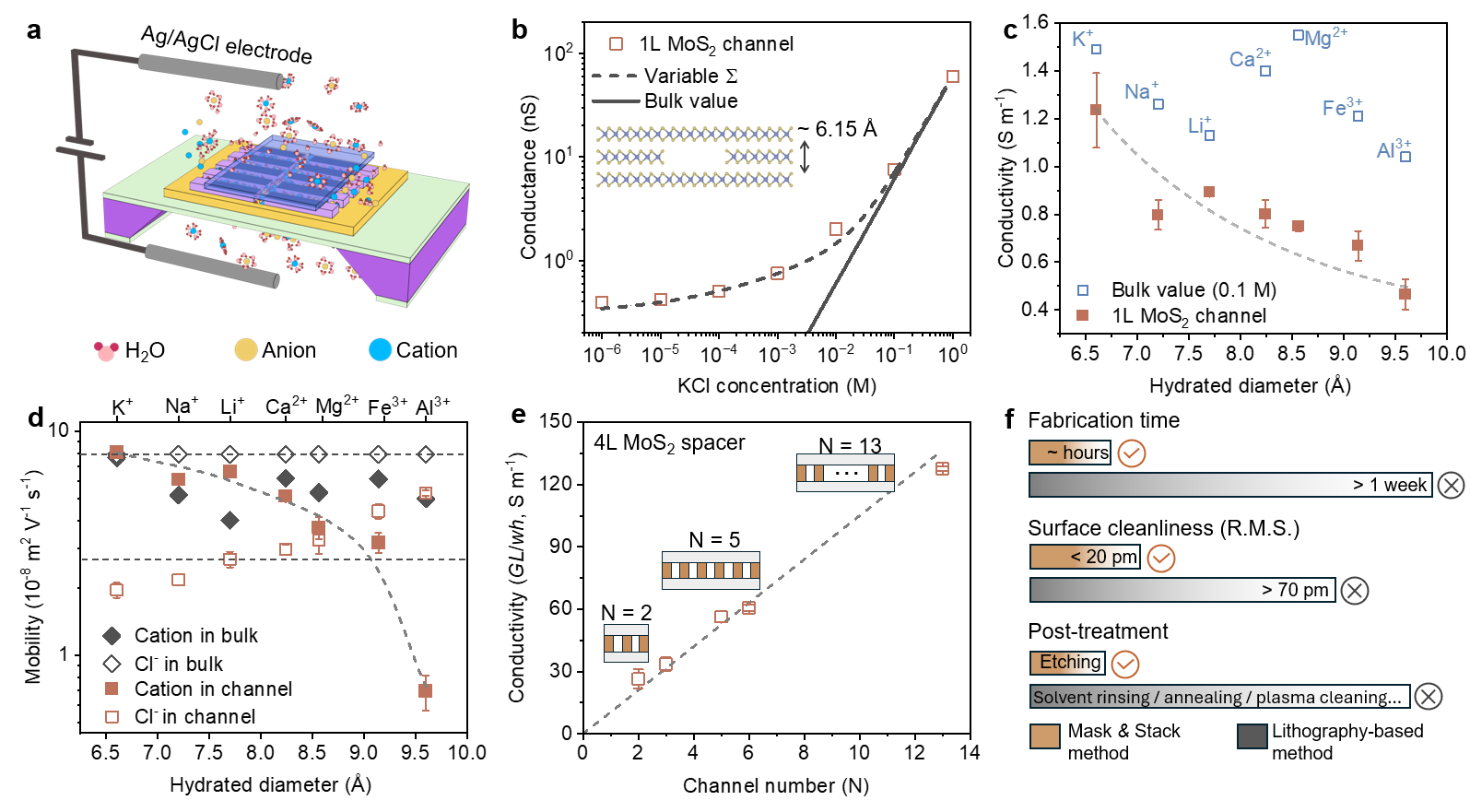}
    \caption{\textbf{Ionic transport through monolayer MoS$_2$ nanochannel}. \textbf{a}, Schematic illustration of the ionic transport measurement setup. \textbf{b}, Conductance for the nanochannel device with 1L MoS$_2$ as the spacer. \textbf{c}, Conductivity of various salt solutions (0.1 M) through nanochannels. Blue open squares denote bulk conductivity of the corresponding salts. \textbf{d}, Effective ion mobility inside the nanochannel (squares) compared to bulk values (diamonds) from literature\cite{nightingale1959phenomenological,haynes2016crc}. \textbf{e}, Ionic conductivity of 4L MoS$_2$ nanochannels as a function of channel number (N). Insets: schematics of devices containing 2, 5, and 13 parallel channels. \textbf{f}, Comparison between  Mask $\&$ Stack method (yellow) and conventional lithography-based methods (gray). }
    \label{Iontrans}
\end{figure}

Drift–diffusion measurements under a 10-fold concentration gradient ($\Delta$C=10) reveal clear ionic selectivity through the channel. Under concentration gradient, we measure diffusion current at zero voltage for different salt solutions and compare the mobilities of the measured ions through monolayer MoS$_2$ nanochannel, as shown in \textbf{Figure 4d}. More details on the drift-diffusion experiments and the analysis can be found in SI notes and \textbf{Figure S9}. The mobility of Cl$^{-}$ under confinement is reduced by a factor of 3 compared to its bulk value. For cations, they exhibit markedly lower mobilities, with the suppression becoming more significant for larger hydrated species. For example, Al$^{3+}$, the largest cation, shows nearly an order-of-magnitude reduction in mobility, reflecting the strong steric constraints imposed by confinement. These results manifest monolayer MoS$_2$ channel as a precise system for probing nanoscale ion dynamics. 

To further evaluate the stability and reproducibility of nanochannels made using the Mask $\&$ Stack method, we fabricated and measured a series of 4L MoS$_2$ channels with varying number of channels. Here, to suppress surface-dominated ion conductance, measurements were carried out in 1M KCl solution (more details about the experimental results and analysis in \textbf{Supplementarty Information, Ionic transport analysis}). We show in \textbf{Figure 4e} that the normalized conductivity agrees well with theoretically expected values and scales linearly with the number of channels. Such consistency confirms the cleanliness of the channels, with no sign of blockage that obstructs the free passage of ions through the channels. \textbf{Figure S9d} compares the measured conductance of multiple devices, covering wide ranges of spacer thicknesses, to theoretical values derived from their channel dimensions. All data points closely follow the theoretical prediction over three orders of magnitude, demonstrating strong agreement between experimental and designed geometries. These results highlight the high precision and reproducibility enabled by our resist-free dry-transfer assembly protocol.

We benchmarked our Mask $\&$ Stack method against conventional resist-based lithography method in terms of fabrication time, surface cleanliness, and post-treatment procedures, as summarized in \textbf{Figure 4f}. The Mask $\&$ Stack method shrinks the weeks-long fabrication process to just a few hours, with significantly improved cleanliness of the channels. Compatible with a wide range of 2D materials, this method enables heterostructure device design and implementation far beyond nanochannels. One can also easily build channels with asymmetric surfaces (see supplementary notes) to explore transport phenomena that are both at interface and under strong confinement. 

The Mask $\&$ Stack method also brings advantages for the fabrication of other 2D heterostructure devices, such as those for electronic transport measurements or nanophotonic devices\cite{gentile2022electronic,jessen2019lithographic, li2021anisotropic, yang2025nanofabrication}. Its resist-free nature ensures clean interfaces among layers, preserving the pristine 2D surfaces. This would enable \enquote{pattern before encapsulation} for devices whose edges need to be protected, for example, for electronic measurements. Additionally, the high etching resistivity of SiN mask compared to common 2D materials means the same mask could be repeatedly used among several devices, guaranteeing the fabrication precision among devices and improving reproducibility. 

Looking forward, with advances in robotics and automation, automated systems for 2D material transfer are rapidly emerging\cite{zhao2025automated,liu2025mass,masubuchi2018autonomous}. Integrating the Mask $\&$ Stack method with robotic transfer systems could transform it into an automated, high-throughput manufacturing pipeline for nanofluidic, nanoelectronic, or nanophotonic devices, minimizing device variability and enabling statistically robust studies across large device spaces.

\section{Conclusion}
In summary, we introduce the Mask $\&$ Stack method for rapidly assembling ultraclean vdW nanochannels. This method eliminates polymer resist contamination and enables reliable fabrication of a wide range of 2D materials heterostructure. Combining speed, quality, and reliability, the Mask $\&$ Stack method is readily adaptable to high-throughput workflows, paving the way for automated, scalable production of nanoscale devices for energy, molecular transport, and precision sensing applications.

\section{METHODS}
\subsection{SiN spacers stencil fabrication}
We first fabricate the SiN suspended membranes. We used double-side polished SiN/Si/SiN wafers (SiN thickness $\sim$130\,nm, MicroChemicals GmbH ) for the fabrication of SiN masks. The fabrication process follows standard photolithography and RIE procedures, as shown in \textbf{ Figure S1a}. Briefly, photolithography was first performed to define a big opening (with size of  800\,$\mu$m~$\times$~800\,$\mu$m) on the SiN substrate, followed by RIE to selectively remove the SiN layer (Step 1). The exposed silicon was then etched in a 30\% KOH solution at 80°C for 10 hours, resulting in suspended SiN membrane on the front side (Step 2). The size of the resulting suspended SiN membranes is approximately 100\,$\mu$m~$\times$~100\,$\mu$m on the front side after KOH etching. A single 4-inch wafer could be made into 69 1-cm coupons after fabrication, suitable for high-throughput and wafer-scale device production (\textbf{Figure S1b}). 

The suspended SiN membrane could be used for both the fabrication of reusable SiN spacer stencil and as the substrate for nanochannel assembly and ionic transport measurements. To make the SiN stencil mask for nanochannels, a few up to tens of parallel lines are designed with desired line width and period, typically 150 nm width and 1 $\mu$m period, are patterned using EBL and RIE (Step 3). Resist CSAR 62-009 (Allresist GmbH)was used, which is more resistive to etching than the commonly used PMMA, while maintaining high resolution patterning. The CSAR 62-009 e-beam resist was spin-coated with 1000\,rpm, followed by a baking step at 170°C for 5 minutes. EBL was performed with the dose of 120\,$\mu$C/cm$^2$ and 25\,kV acceleration voltage. The exposed patterns were developed for 1 minute using standard CSAR developer. Then, the patterned SiN was etched using RIE for 18\,s + 18\,s + 10\,s to minimize resist crosslinking. The samples were further treated with UV-O$_3$ for 10 minutes before resist removal. After that, the resist was removed by soaking in hot remover solution (70°C) for 2 hours, followed by immersion in piranha solution for several minutes. Finally the SiN stencils were soaked in hot deionized water for 2 hours to remove any remaining chemicals before drying up. Before using the SiN mask for device fabrication, an O$_2$/Ar plasma cleaning step is usually performed to remove hydrocarbons on the surfaces. We used Vistec electron beam lithography system to make the masks which allows 4-inch wafer-scale patterning at once, with spatial resolution approaching  10 nm, as shown in \textbf{Figure S2}. For SiN/Si/SiN to serve as nanochannels substrates, a slit window (3\,$\mu$m~$\times$~20\,$\mu$m) is patterned on the suspended SiN membrane by repeating the same photolithography and RIE procedures. The SiN/Si/SiN substrate will then separate two liquid reservoirs for ionic transport measurements.

\subsection{Dry transfer method for vdW assembly}
The sandwiched nanochannel structure was assembled using a PDMS-assisted dry transfer method. Before the dry transfer process, the PDMS stamp was thoroughly cleaned by sequential rinsing in acetone, isopropanol (IPA), and hexane. As shown in \textbf{Figure 1a},  the spacer crystal was mechanically exfoliated onto a PDMS stamp and then transferred onto the patterned SiN stencil mask at 50°C in step 1. RIE was then performed from the backside to define the spacer pattern. In step 2, another freshly exfoliated crystal was exfoliated onto a PDMS stamp to serve as the top crystal. The top crystal was chosen to be larger than the spacer to ensure full coverage during the pick-up process. The pick up temperature is around 150°C, which helps to expel PDMS residues at the interface between the top crystal and the spacer, promoting clean and strong vdW adhesion\cite{jain2018minimizing}. In step 3, the bottom crystal was first exfoliated onto a Si/SiO$_2$ substrate. Then, the top/spacer PDMS stack is used to align and pick up the bottom crystal at 180°C to complete the tri-layer assembly. In Step 4, the assembled stack was subsequently released onto a suspended SiN window substrate at around 100°C, completing the encapsulated nanochannel device. 

We truncated the encapsulated channels just before measurements to protect the channels from hydrocarbon contamination. For ionic transport measurements, we opened both ends of the channels to allow mass transport, as shown in the red arrow in \textbf{Figure 1b}. To define the front entrance, a gold mask (3/50\,nm Cr/Au) is usually made via photolithography and e-beam evaporation, serving as an etch mask for subsequent RIE. This process allows precise control over the number and length of the nanochannels. Alternatively, another crystal can be dry transferred as an etching mask on top of the nanochannel. For example, hBN can serve as etching mask for graphite or vise versa, because of different etching conditions for them. For the back entrance, two strategies were employed: (1) aligning the edge of the bottom crystal with the edge of the SiN slit so that the channels are naturally opened through the SiN slit, or (2) backside RIE to etch through the bottom crystal. This applies to the situation where the top and bottom are different crystals or the situation where the top crystals are much thicker than the bottom one if they are the same material. The Mask and Stack method was used for the fabrication of all nanochannel devices carried out in the clean room. A detailed video demonstration of the fabrication process is included to facilitate replication of the procedure, see \textbf{supplementary video}.

\subsection{AFM measurements}
AFM was employed to evaluate the surface cleanliness and roughness of the fabricated nanochannel devices. AFM measurements were performed using Asylum research Cypher ES system (Oxford Instrument). To characterize the spacer surface, the patterned spacers were picked up by a top crystal that was supported on a PDMS stamp. RMS roughness values were extracted from $0.5\,\mu\text{m} \times 0.5\,\mu\text{m}$ scan areas using Gwyddion software. For the capillary filling experiments where we examined the height profiles of the channels before and after water filling, we performed AFM in air and deionized water using Cypher ES system in contact mode using the same cantilever (ScanAsyst-Fluid Plus) and the same set-point force. The height profiles for empty and filled channels were compared in the same scanning area.

Note that the AFM characterization of our spacers was performed directly on top of a thick top crystal (typically graphite, hBN or MoS\textsubscript{2}), which yields intrinsically low roughness because the crystal surface is atomically flat. In contrast, previously reported EBL-fabricated spacers were evaluated on SiO\textsubscript{2} substrates, which could increase apparent surface roughness. To provide a fair baseline, we measured freshly exfoliated few-layer graphene (FLG) on the same SiO\textsubscript{2} substrate (\textbf{Figure} \textbf{S5}\textbf{a} and \textbf{b}). Its RMS roughness ($\sim$145 pm) is substantially lower than that of the PMMA-assisted EBL spacer ($\sim$216 pm), indicating that the increased roughness arises from PMMA residues in addition to the SiO\textsubscript{2} substrate itself. 

\subsection{Raman spectroscopy }
Raman measurements were performed using an Invia Raman microscope (Renishaw, UK) with a 532 nm continuous-wave laser. The laser beam was focused onto the graphene channel using a 100× objective lens with a numerical aperture of 0.9, and the scattered light was dispersed by a 1800 grooves per mm holographic grating. This configuration enabled reliable characterization of disorder, strain, and doping. Spectrometer calibration was carried out by recording the first-order Raman peak of a silicon reference at 520.5 \(\text{cm}^{-1}\).

\subsection{Ionic transport measurements}
For ionic transport experiments,  nanochannel device was mounted between a symmetric H-type electrochemical cell and sealed with an O-ring to ensure that the nanochannels provided the only pathway between the two reservoirs. Two homemade Ag/AgCl electrodes were used as working electrodes to record ionic currents. To prepare the Ag/AgCl electrode, Ag wire (0.8 mm diameter) was first polished and then coated with a layer of AgCl by electroplating using a Keithley 2461 source meter. In this procedure, the Ag wire served as the anode, a Pt plate as the cathode, and 1 M KCl solution as the electrolyte. A constant current of 1 mA was applied to the Ag wire for 30 min within a voltage window of –1 to –2 V. The resulting Ag/AgCl electrodes, identifiable by their dark brown color, were rinsed with deionized water and stored in 1 M KCl solution. The potential difference between two such electrodes in 1 M KCl was verified to be less than 5 mV. Commercial Ag/AgCl (3 M KCl) reference electrodes were used during drift–diffusion tests to eliminate the Nernst potential difference. Before measurements, the cell was thoroughly rinsed with IPA to pre-wet the channel sample, then with 100 mL of DI water to ensure complete IPA removal. Finally, between measurements with electrolyte of different concentrations, the cell was flushed with large volume of the test electrolyte until a stable open circuit current was obtained. All tests are performed from low to high concentration. The I-V curve was performed using a Keithley 2636B source meter which is controlled by a custom LabVIEW program. 

\newpage
\setcounter{figure}{0}
\renewcommand{\thefigure}{S\arabic{figure}}
\section{Supplementary Notes}
\subsection{Raman analysis}
\textbf{Figure 8a }and\textbf{ b} show the optical and AFM images of the nanochannel device used for Raman measurements, with the white dashed line indicating the position of the embedded MLG in the device. Raman mapping reveals a highly uniform distribution of both G and 2D peak parameters across the nanochannel region, indicating clean interfaces between MLG and the surrounding hBN layers. As shown in \textbf{\textbf{Figure 8c}}, the 2D peak intensity exhibits periodic modulations that match the nanochannel spacing, suggesting local strain variations induced by the alternating suspended and supported channel geometry. Furthermore, the $I_{2D}/I_G$ intensity ratio remains high (\(>\)3) across the graphene-embedded area, indicating negligible doping effects and minimal polymer contamination, as shown in \textbf{Figure 8d }. 

To further investigate the interplay between strain and doping, we plotted the 2D versus G peak positions, with color-coded  $I_{2D}/I_G$ ratios (\textbf{Figure 8e}).  Most data points follow a linear correlation consistent with uniaxial strain (data points on the left branch), while a subset (datapoints on the right branch) deviates from this trend, indicative of local doping effects\cite{lee2012optical}. The observed G peak upshift corresponds to a carrier density of approximately  $6 \times 10^{12} \,\text{cm}^{-2}$, equivalent to a Fermi level shift of $E_F \approx 0.3 \,\text{eV}$. This behavior suggests spatially localized doping or disorder, likely arising from strain-induced charge inhomogeneities. These findings are consistent with earlier reports on the graphene/hBN heterostructure, where strain–doping decoupling has been widely used to evaluate interfacial quality\cite{lee2012optical}. 

Normalized Raman spectra of the embeded MLG obtained from the channel region (blue) and encapsulated region (red) are shown in \textbf{Figure 8f}. In addition to the characteristic graphene G ( $\sim$1580 \(\text{cm}^{-1}\)) and 2D ( $\sim$2680 \(\text{cm}^{-1}\)) bands, a peak below 1400 \(\text{cm}^{-1}\) arises from the $E_{2g}$ vibrational mode of the top hBN \cite{son2017graphene}. Both spectra exhibit negligible D peak intensity, confirming the high quality of MLG. 

\subsection{Ion transport analysis}
To extract the surface charge density  \(\Theta\)  and individual ionic mobilities, we followed the procedure report by \textit{Esfandiar et al}\cite{esfandiar2017size}. The conductance \(G\) was analyzed using a variable surface charge model. To obtain the absolute ionic mobilities, we first measured the ionic conductivity of various electrolyte solutions through the nanochannel (as shown in \textbf{Figure 4c}). The conductivity \(\sigma\) is then approximated by: \begin{equation}
    \sigma \approx F(c_+\mu^+ + c_-\mu^-)
    \label{eq:placeholder}
\end{equation}
where \(c_+\) and \(c_-\) are the concentrations of cations and anions, respectively. 

For drift–diffusion experiments, as shown in \textbf{Figure S9a}, a positive diffusion current at zero voltage indicates that cations (e.g., K$^{+}$) diffuse faster than anions (Cl$^{-}$) through the nanochannel, whereas a negative current reflects slower cation transport, as observed for Al$^{3+}$. Moreover, the resulting zero-current potential \(E_m\), was analyzed using the Henderson equation to extract the ionic mobility ratio,  \(\mu^+/\mu^-\).  We extracted the effective mobility ratio between cations and anions for all the tested salts, as shown in \textbf{Figure S9b}. This ratio decreases by nearly an order of magnitude from K$^{+}$ to Al$^{3+}$, correlating with increasing hydrated ion diameter. The strong size dependence confirms the presence of steric exclusion effects arising from the 1L MoS$_2$ channel. We then calculated the mobilities of cations and anions by combining the measured conductivity with the extracted mobility ratios, as shown in \textbf{Figure 4d} in the main text.

To verify the uniformity of our fabrication process, we fabricated nanochannels with the same height (4L MoS$_2$ spacers) but varying numbers, lengths, and widths, and then compared their normalized conductance. 
\begin{equation}
   G = N \cdot \sigma \cdot \frac{wh}{L}
\end{equation}
where \(N\), \(\sigma\), \(w\), \(h\) and \(L\) are the channel number, ionic conductivity of 1M KCl, width, height and length of channels, respectively. By normalizing the geometric factor  \({wh}/{L}\), we evaluated the dependence of conductance on channel number. As shown in \textbf{Figure 4e}, the normalized conductance  \(GL/wh\) scales linearly with \(N\), in agreement with the theoretical values. This confirms the high yield and device-to-device consistency of our fabrication process with no sign of blockage of any channels. Moreover, as shown in \textbf{Figure S9c}, the conductance of the same 4L MoS$_2$ channel remains stable over at least 7 days under ambient conditions, showing good structural robustness and environmental stability. Note that we do observe the gradual blockage of channels over extended period (after several weeks) of ionic transport measurements, despite the salt solutions were filtered before filling the cell.  This may be due to dissolved airborne contaminants that accumulate and lead to degradation of the device performance over a long time. \textbf{Figure S9d} compares the measured conductance of multiple devices, covering wide ranges of spacer thicknesses, to theoretical values derived from their channel dimensions. All data points closely follow the theoretical prediction over three orders of magnitude, demonstrating good agreement between experimental and designed geometries. These results highlight the high precision and reproducibility enabled by our Mask $\&$ Stack method.

\subsection{Heterostructure Channels with Interfacial Asymmetry }
The Mask $\&$ Stack method also enables the fabrication of heterostructure channels with precisely tunable interfacial asymmetry. As a demonstration, we assembled channels consisting of hBN (top), MoS$_2$ (spacer), and graphene (bottom) (\textbf{Figure S10}). This heterostructure design creates deliberate interfacial asymmetry for nanoconfinement effects\cite{xiao2016biomimetic}, enabling experimental exploration of ion transport under asymmetric surface potentials\cite{qiao2023ion}, structure of nanoconfined water and molecules near heterogeneous boundaries\cite{fumagalli2018anomalously,qiao2003ion}, or even electrochemical gating effects\cite{laucirica2023electrochemically}. Moreover, by tuning the composition and stacking order of the constituent 2D layers, one can modulate the dielectric environment and interfacial interactions within the channel, offering unprecedented control over ion–solid and water–solid coupling at the atomic scale\cite{aluru2023fluids}. By combining stencil-defined architectures with the vast library of 2D materials, one can also envisage the construction of programmable nanofluidic platforms for energy harvesting, molecular sensing, and quantum-confined transport studies.

\newpage
\begin{figure}
    \centering
    \includegraphics[width=1\linewidth]{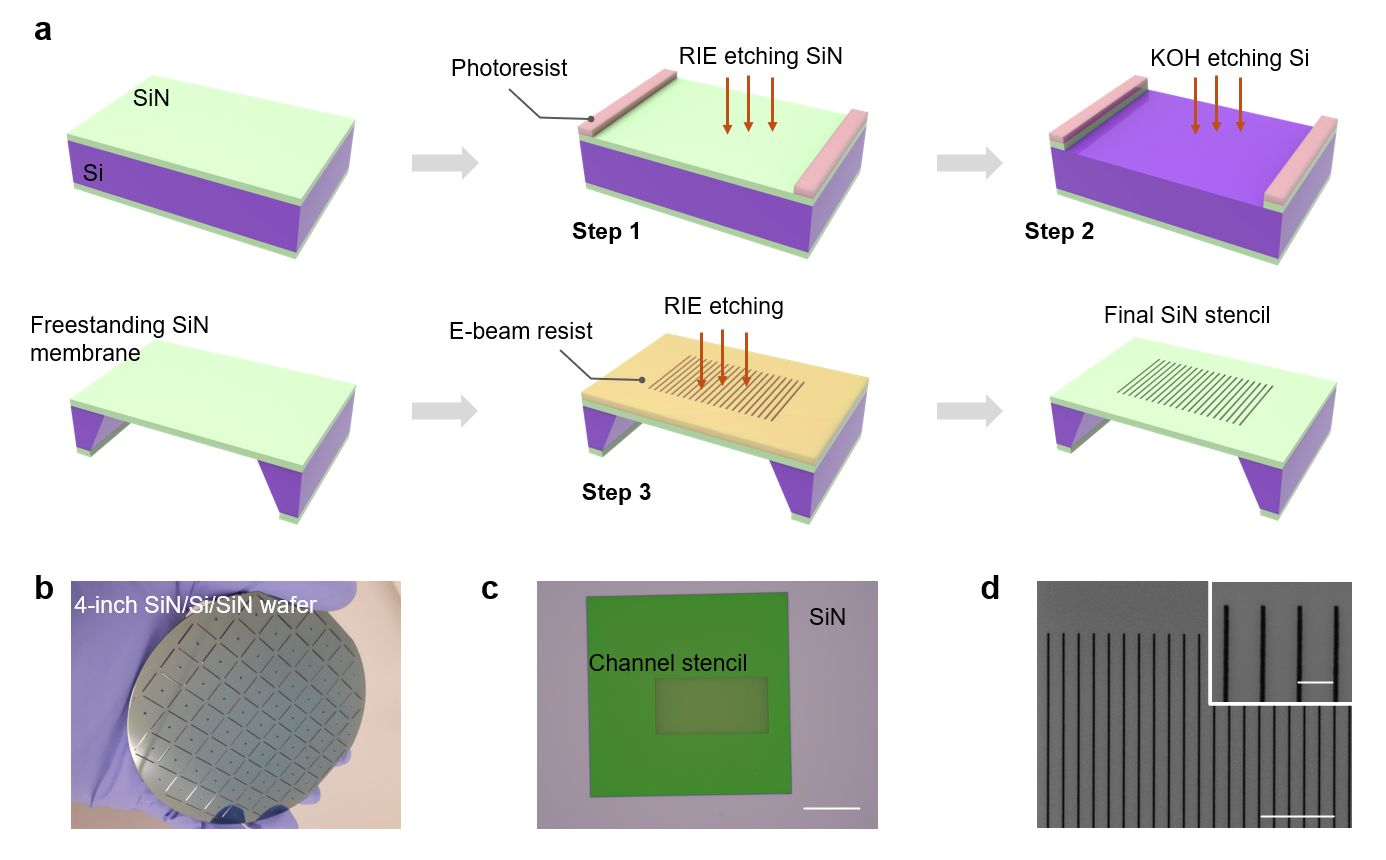}
    \caption{\textbf{Fabrication and characterization of the SiN stencil.} \textbf{a}, Schematic illustration of the fabrication process. A SiN/Si/SiN wafer was processed by photolithography and RIE to define the a large SiN window (step 1). The exposed Si is then selectively removed by KOH etching (step 2) to form a freestanding SiN membrane. EBL and RIE are then used to pattern parallel lines into the freestanding SiN membrane (step 3), yielding the final SiN stencil for nanochannels. \textbf{b}, Photograph of a 4-inch SiN/Si/SiN wafer patterned with 69 SiN stencil masks, demonstrating wafer-scale production and reproducibility of the process. Wafer-scale production enables its upscale using automated assembly workflows and potential integration with CMOS architectures. \textbf{c}, Optical image of a SiN stencil with patterned nanochannels. \textbf{d}, SEM images of SiN stencil. Inset: zoom in version of the channel area. Scale bar: \SI{30}{\micro\meter} in c,  \SI{10}{\micro\meter} in d, and \SI{2}{\micro\meter} in the inset of d.}
    \label{Fabprocess}
\end{figure}

\newpage
\begin{figure}
    \centering
    \includegraphics[width=1\linewidth]{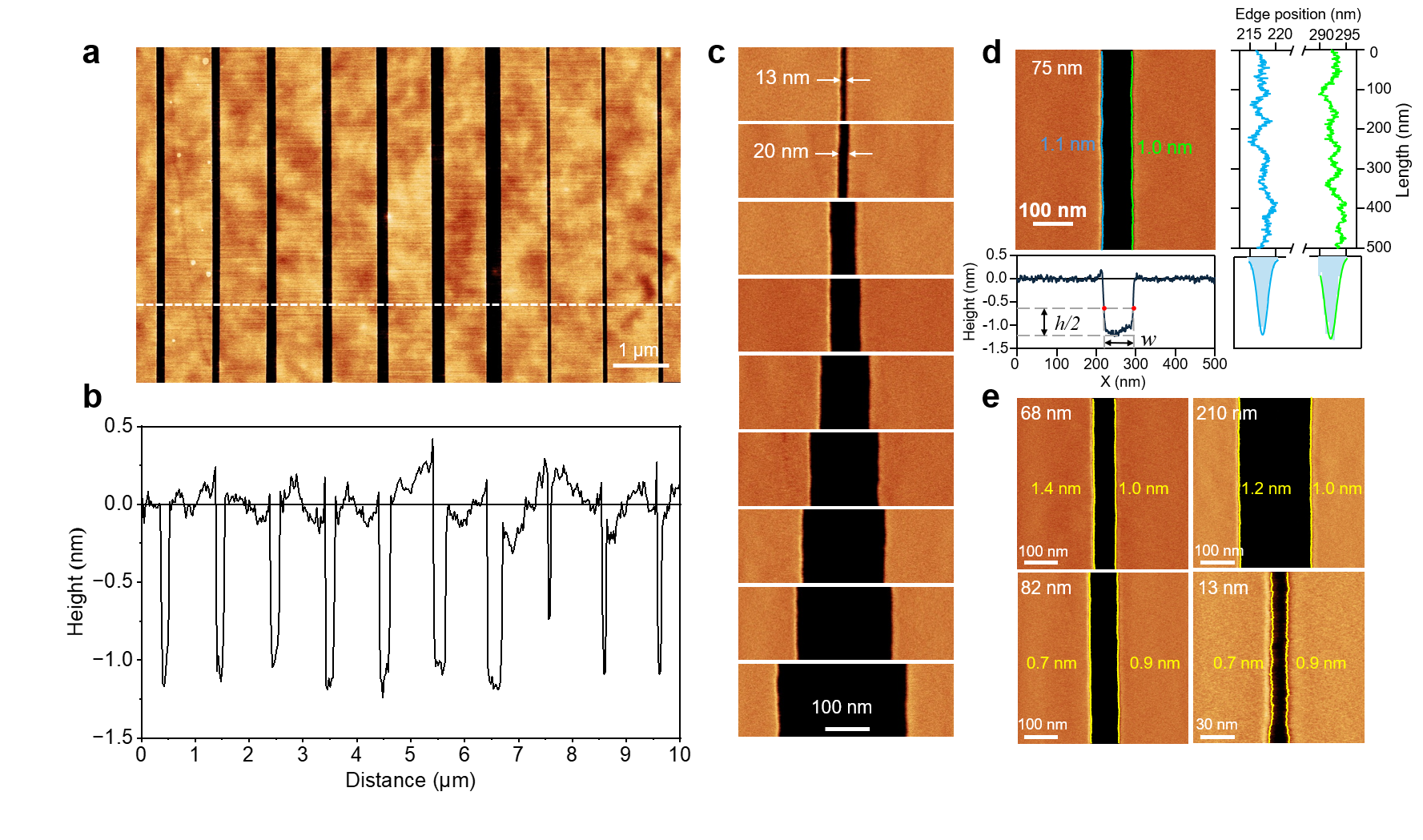}
    \caption{\textbf{AFM images of MoS\textsubscript{2} spacer etched}. \textbf{a}, Bi-layer MoS\textsubscript{2} spacers with different channel width design. \textbf{b}, Height profile across the white dashed line in \textbf{a}, the uneven background comes from the substrate underneath. \textbf{c}, Snapshot of individual MoS\textsubscript{2} channels extracted from a, with the narrowest channel of only around 10 nm. This would enable the construction of 1D nanochannels with geometry approach that of nanotubes. This 13-nm-wide channel is around 20 $\mu$m long, with aspect ratio reaching 2000. Even thinner spacers are prone to collapse because of the vdW forces. \textbf{d}, AFM topograph of a 75-nm-wide MoS\textsubscript{2} channel with smooth, well-defined edges (left top panel).  Representative line-scan profile illustrating the edge-position extraction using the mid-height criterion (left bottom panel). For each scan line, the height values on the two flat plateaus—the spacer region and the bottom MoS\textsubscript{2} surface—are first identified, and their mean values are extracted. The edge position is then defined as the precise lateral coordinate at which the profile crosses the midpoint height, the 50\% level of the step between the two plateaus. This crossing is determined by interpolating between adjacent pixels, providing sub-pixel accuracy and avoiding discretization artifacts. Positions of both channel edges along the full channel length and their corresponding distributions; the line-edge roughness (LER) is defined as the standard deviation (right panel). \textbf{e}, Gallery of channels on the same sample with extracted edge positions overlaid and corresponding LER values. The LER is independent of pixel size (0.98 – 0.58 nm), and the extracted correlation lengths (25 – 93 nm) indicate slowly varying, highly coherent edge smoothness\cite{chou2008improved}. }
    \label{fig:placeholder}
\end{figure}

\newpage
\begin{figure}
    \centering
    \includegraphics[width=1\linewidth]{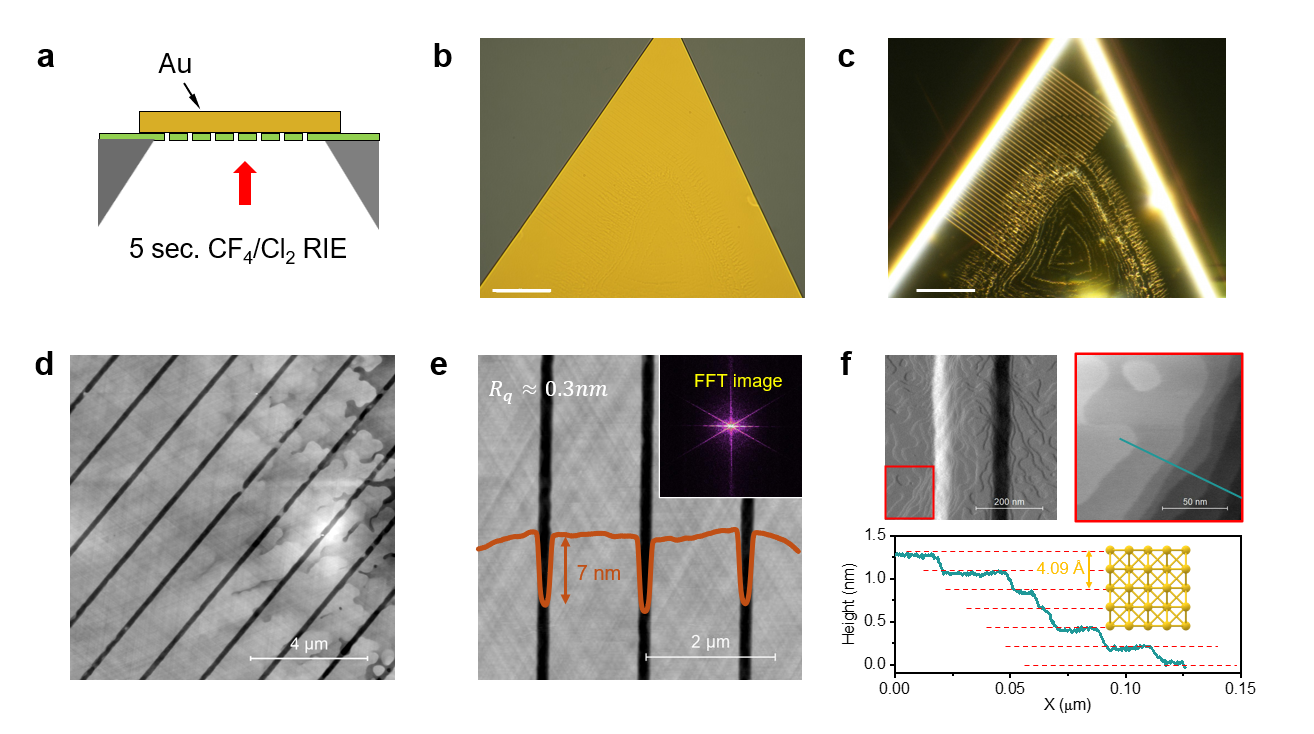}
    \caption{\textbf{Stencil etching of crystal Au.} \textbf{a}, Schematic illustration of the Au etching process using a SiN stencil. The Au is etched by RIE to define nanochannels on 2D single crystalline Au. \textbf{b}, Optical image of the Au surface after stencil etching.  \textbf{c}, Corresponding dark-field optical image highlighting the etched parallel linesnear the left edge. The triangular patterns in the center of the crystal are the growth patterns of Au. \textbf{d}, AFM image of etched Au channel. \textbf{e}, Zoomed-in AFM image from (d) with representative height profile indicating $\sim$7 nm channel depth. Inset: fast Fourier transform (FFT) image, showing periodicity of Au crystal. The root mean square (RMS) surface roughness is $\sim$0.3 nm. \textbf{f}, The top-left panel shows the AFM deflection signal of one channel in \textbf{e}. The top-right panel presents a zoomed-in AFM image of the red box region from top left panel. The green line indicates the position where the height profile in bottom panel was extracted, which shows atomic-scale steps with a periodic height of $\sim$4.09 Å, consistent with two atomic steps of Au (111) ($\sim$4.08 Å)\cite{sandy1991structure}. The presence of atomically flat terraces within the channel confirms that the stencil etching process preserves both the crystallinity and atomic scale cleanliness. Scale bars: \SI{20}{\micro\meter} in b and c. }
    \label{Auspacer}
\end{figure}

\newpage
\begin{figure}
    \centering
    \includegraphics[width=1\linewidth]{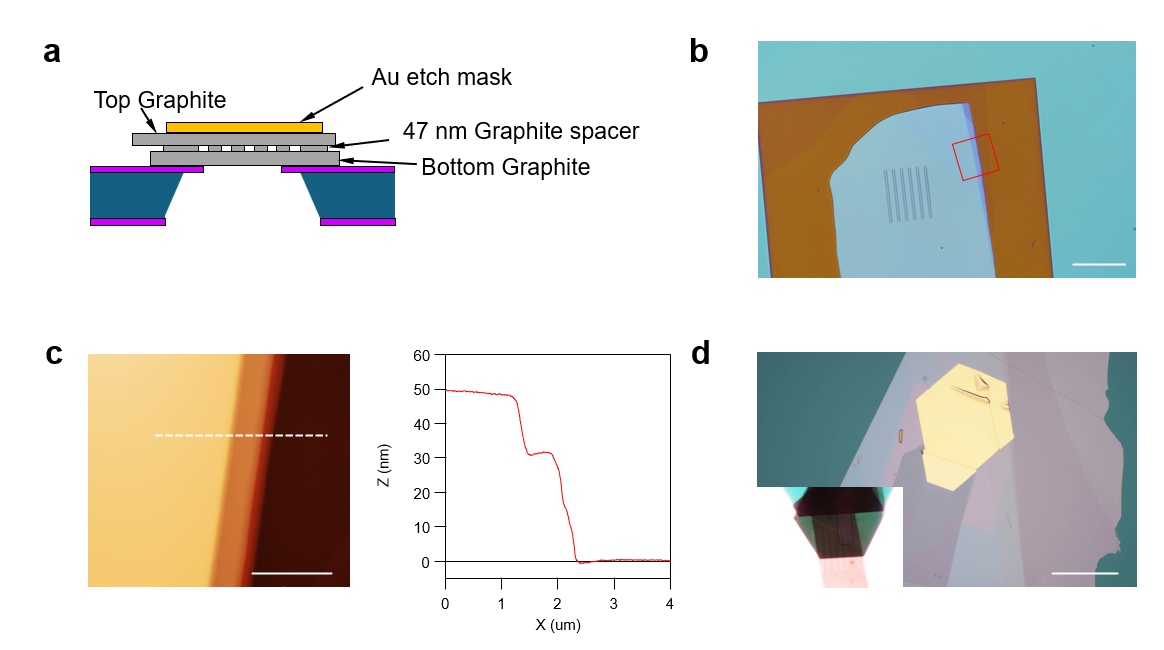}
    \caption{\textbf{Fabrication of the nanochannel with 47 nm thick graphite spacer. } \textbf{a}, Schematic cross-section of the nanochannel structure, consisting of a bottom graphite layer, a 47 nm thick graphite spacer, a top graphite cover, and a gold etch mask. \textbf{b}, Optical image of the graphite spacer after etching. \textbf{c}, AFM image and corresponding height profile of the graphite spacer. \textbf{d},Optical image of the fabricated nanochannel device before opening the channels using RIE. Here we use the sharp edge of a dry-transferred Au crystal as the etching mask to open the channels. The Au etch mask (yellow part) protects the channel region and determines the channel length. Inset: Backside optical image in transmission mode through the SiN window, highlighting the well-defined channel geometry. Scale bars: \SI{20}{\micro\meter} in b,  \SI{20}{\micro\meter} in c and \SI{50}{\micro\meter} c.  
}
    \label{fig:placeholder2}
\end{figure}

\newpage
\begin{figure}
    \centering
    \includegraphics[width=1\linewidth]{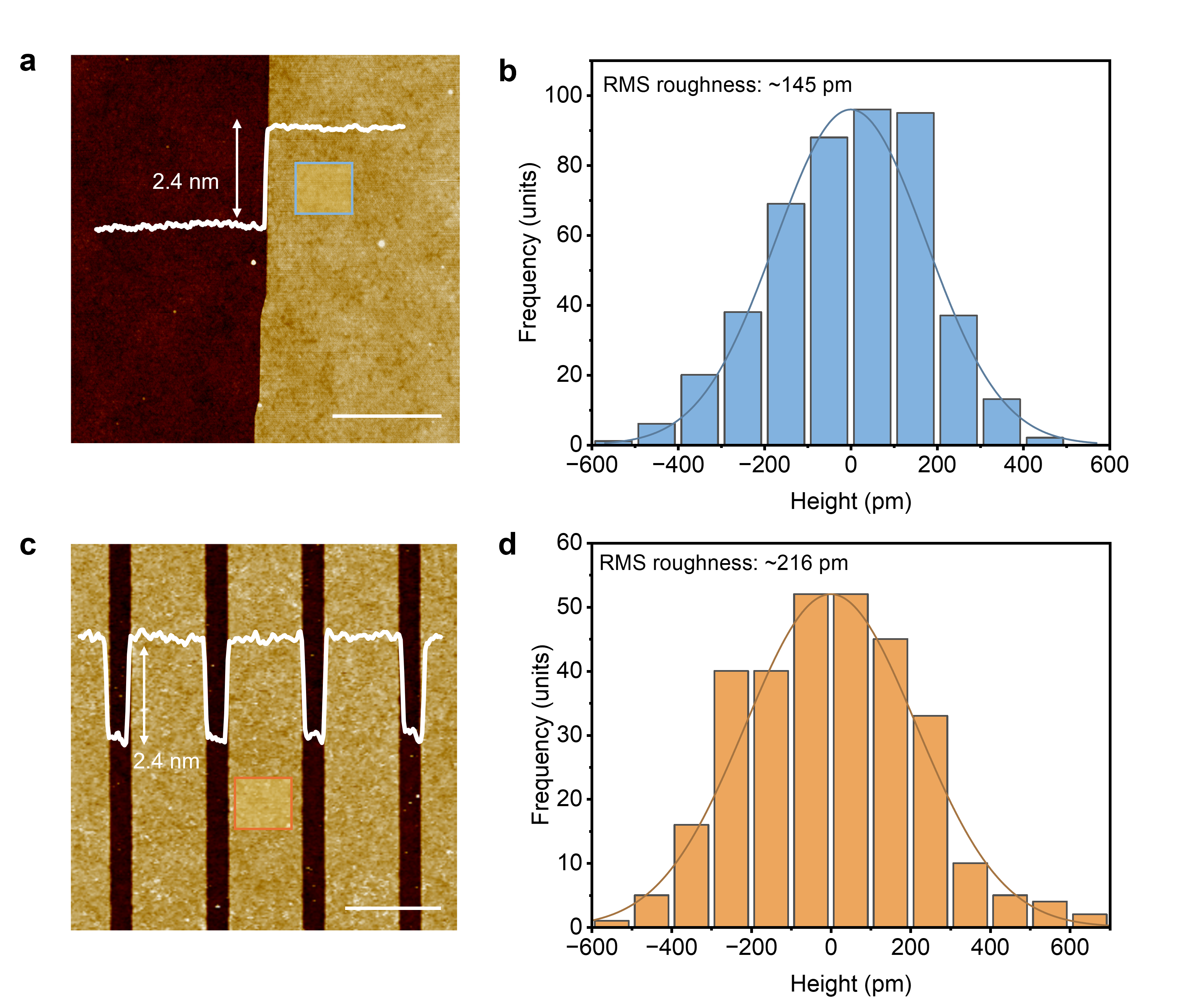}
    \caption{\textbf{AFM analysis of a few graphene spacer fabricated using EBL and the comparison to freshly exfoliated graphene.} \textbf{a}, AFM image and corresponding height profile of freshly exfoliated few-layer graphene (FLG) on an oxidized silicon (SiO\textsubscript{2}/Si) substrate, showing thickness of 2.4 nm (7 layers). \textbf{b}, Height histogram from the boxed region in \textbf{a.}, giving an RMS roughness of $\sim$145 pm, which reflects the combined intrinsic flatness of clean FLG and the underlying SiO\textsubscript{2} substrate. \textbf{c}, AFM image and corresponding height profile of the EBL-defined graphene spacer, indicating the same thickness of $\sim$2.4 nm. The spacers appear clean as it underwent prolonged solvent rinsing with sonication, followed thermal annealing to remove residual PMMA. \textbf{d}, Height histogram from the boxed region in (c), showing an RMS roughness of $\sim$216 pm. Scale bars: \SI{1}{\micro\meter} in \textbf{a} and \textbf{b.}}
    \label{AFM_EBLspacer}
\end{figure}

\newpage
\begin{figure}
    \centering
    \includegraphics[width=1
\linewidth]{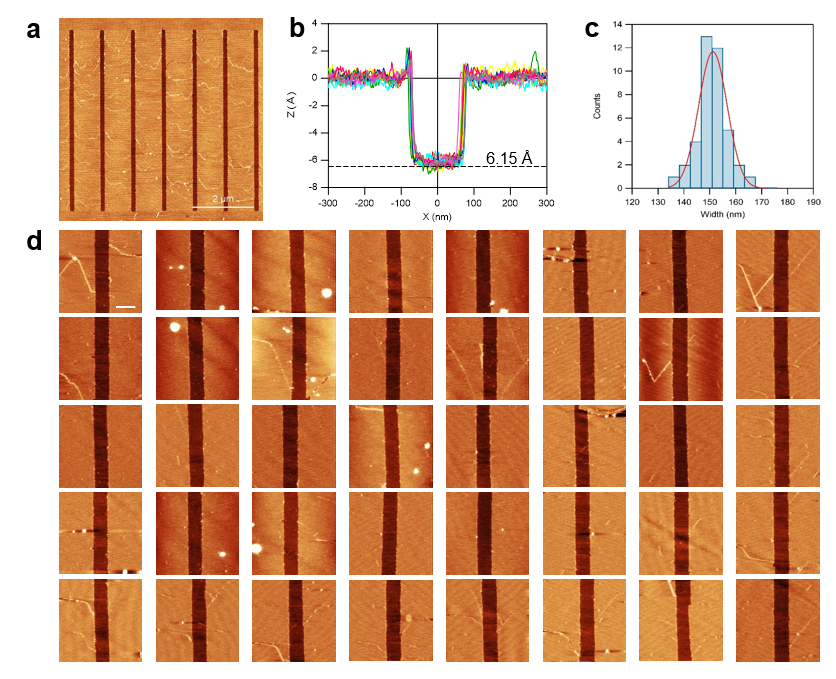}
    \caption{\textbf{AFM characterization of different monolayer MoS\textsubscript{2 }spacers made using the same stencil mask.} \textbf{a}, Large area AFM scan of the monolayer MoS\textsubscript{2 }spacer on bottom MoS\textsubscript{2 } showing uniform topography and clean channel edges. Wrinkles appear sometimes because of strain during dry transfer. \textbf{b}, Cross-sectional profiles extracted from ten different channels shown in panel \textbf{d}, revealing a step height of $\sim$6.15 Å, consistent with interlayer spacing of  2H-MoS\textsubscript{2 }crystal.\textbf{c}, Histogram of channel widths from AFM images, fitted with a Gaussian distribution centered at $\sim$150 nm, consistent with the width of the SiN mask used. \textbf{d}, AFM images of 40 individual nanochannels etched using the same SiN mask, demonstrating high uniformity in dimension of the nanochannel. Scale bar: 200 nm for all panels in \textbf{e.} }
    \label{Widthdistribution}
\end{figure}

\newpage
\begin{figure}
    \centering
    \includegraphics[width=1\linewidth]{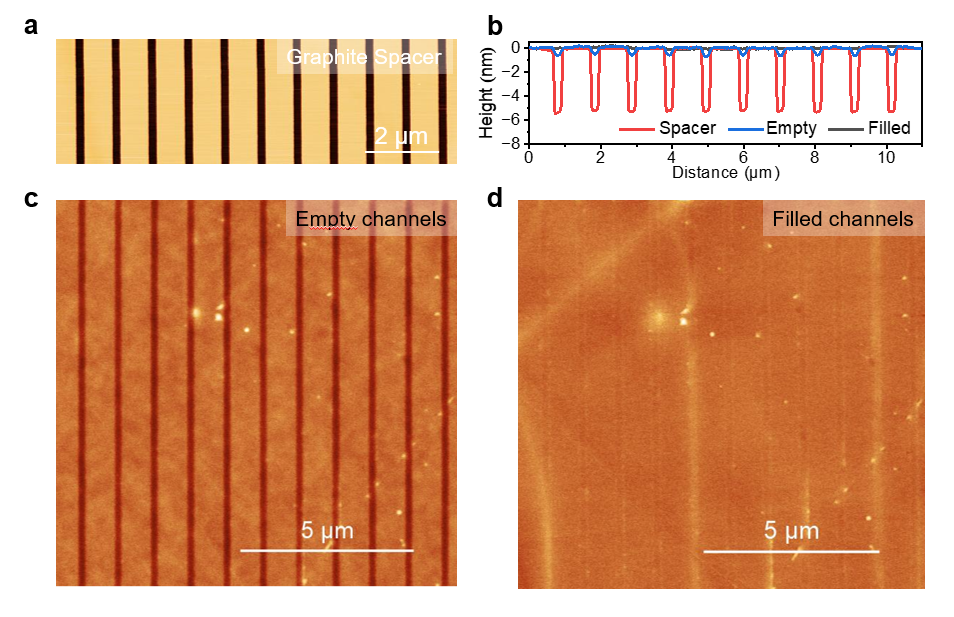}
    \caption{\textbf{Water filling into nanochannels characterized by AFM} \textbf{a}, AFM image of the graphite spacers. \textbf{b}, Height profiles across spacers and channels with and without water. \textbf{c}, AFM image of nanochannels in the dry state (no water filling), showing clear and continuous sagging. \textbf{d}, AFM image of the same channel region after water fills the channels. The sagged channel areas level up to the spacers nearby. This is the same device as shown in \textbf{Figure 2h-j}, and the color scale range is kept identical among all AFM images. }
    \label{WaterfillingS}
\end{figure}

\newpage
\begin{figure}
    \centering
    \includegraphics[width=1\linewidth]{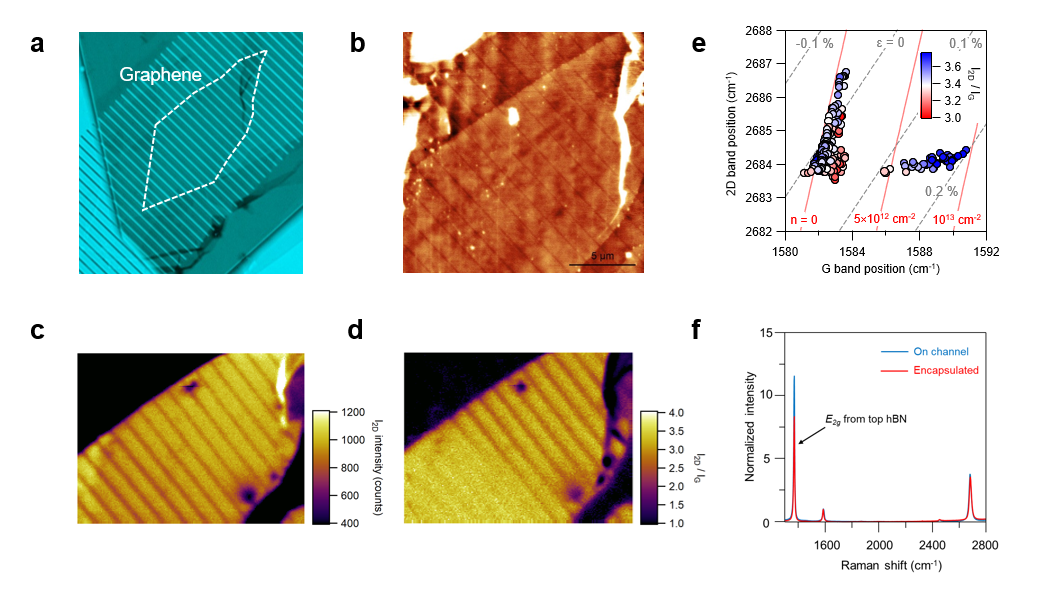}
    \caption{\textbf{Raman characterization of MLG-embedded hBN nanochannels.} \textbf{a}, Optical images (green filter applied) of the nanochannels with the embedded MLG outlined by the white dashed line. \textbf{b}, Corresponding AFM image showing the surface morphology and periodic spacer structures. \textbf{c}, Raman mapping of the same region, where the color scale represents the intensity of the 2D peak (\(I_{2D}\)). \textbf{d}, Raman mapping of the \({I_{2D}}/{I_{G}}\) ratio, confirming the high quality of embedded MLG. \textbf{e}, Evolution of the 2D peak position versus G peak position along the line scan indicated by the white dashed line in main text \textbf{Figure 3b}. Each data point corresponds to a single spectrum along the scan direction, with color indicating the intensity ratio \({I_{2D}}/{I_{G}}\). \textbf{f}, Representative Raman spectra acquired from two regions: above channel (blue) and above spacer (red).}
    \label{RamanS1}
\end{figure}

\newpage
\begin{figure}
    \centering
    \includegraphics[width=1\linewidth]{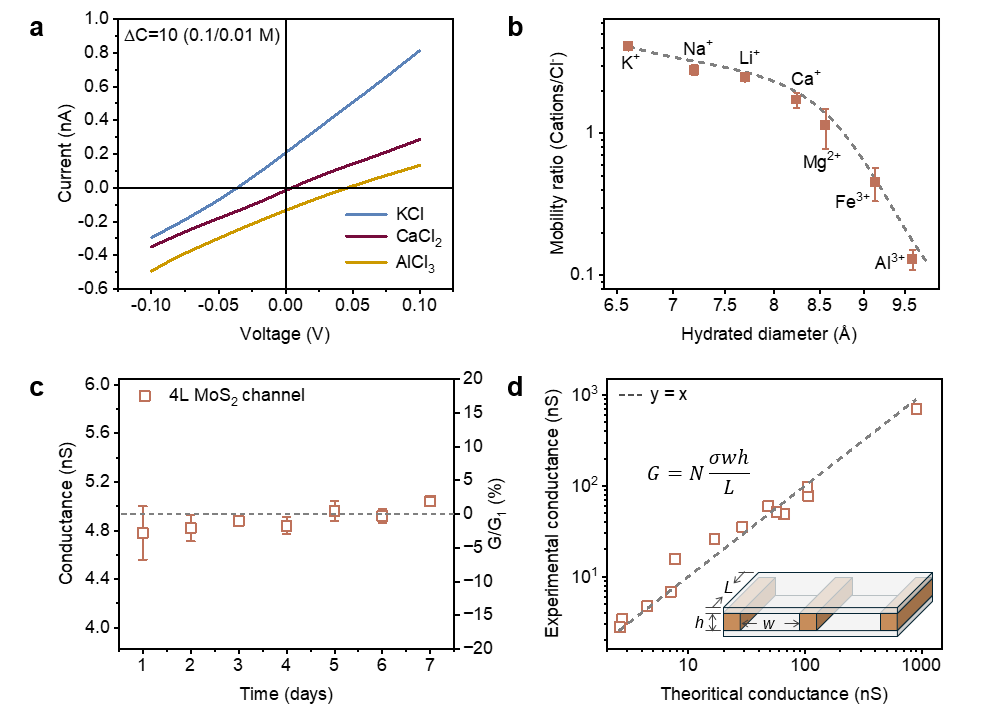}
    \caption{\textbf{Ionic transport behaviors in  monolayer MoS\textsubscript{2} nanochannel.}  \textbf{a}, I-V characteristics for various chloride electrolytes under 10-fold (\(\Delta\) = 10) concentration gradient. \textbf{b}, Mobility ratio \(\mu^+\)/\(\mu^-\) as a function of cations’ hydrated diameter, showing strong size-dependent transport behavior. \textbf{c}, Conductance stability of a 4L MoS$_2$ nanochannel over 7 days of continuous measurement. \textbf{d}, Experimental conductance plotted against theoretical predictions based on channel geometry. Dashed line represents the ideal case y = x.}
    \label{iontransSI
}
\end{figure}

\newpage
\begin{figure}
    \centering
    \includegraphics[width=1\linewidth]{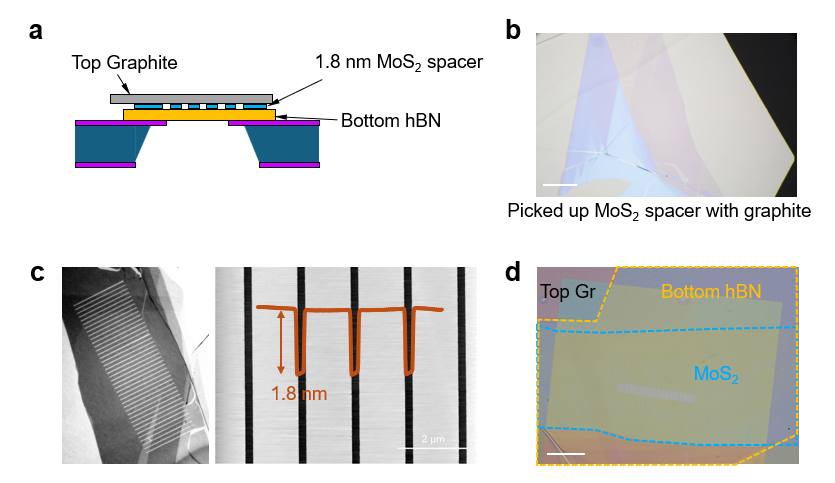}
    \caption{\textbf{Fabrication of nanochannels using heterogeneous vdW crystals.} \textbf{a}, Schematic cross-section of the nanochannel structure, consisting of a bottom hBN, a 1.8 nm MoS\textsubscript{2} spacer, and a top graphite. \textbf{b}, Optical image of the MoS\textsubscript{2} spacer after pickup with a graphite flake. \textbf{c}, The optical MoS\textsubscript{2} spacer image in gray scale highlighting the MoS\textsubscript{2} channel region and corresponding AFM image and height profile of the MoS\textsubscript{2} spacer, confirming a thickness of $\sim$1.8 nm. \textbf{d}, Optical image of the fully assembled device with hBN/MoS\textsubscript{2}/graphite heterostructure. }
    \label{fig:placeholder}
\end{figure}

\clearpage

\bibliography{Main}
\end{document}